\documentstyle[12pt]{article}
\def\baselinestretch{1.5}

\newcommand{\be}{\begin{equation}}
\newcommand{\ee}{\end{equation}}
\newcommand{\ba}{\begin{eqnarray}}
\newcommand{\ea}{\end{eqnarray}}
\newcommand{\baa}{\begin{eqnarray*}}
\newcommand{\eaa}{\end{eqnarray*}}
\newcommand{\bb}{\begin{thebibliography}}
\newcommand{\eb}{\end{thebibliography}}
\newcommand{\ci}[1]{\cite{#1}}
\newcommand{\bi}[1]{\bibitem{#1}}
\newcommand{\lab}[1]{\label{#1}}
\title{
{\bf  Supercritical Pomeron
and  the eikonal represantation \\
of the diffraction processes }}
\author{\bf S.V.Goloskokov, S.P.Kuleshov\\
Bogoliubov Laboratory of Theoretical Physics \\
Joint Institute for Nuclear Research, Dubna }

\begin{document}
\begin{center}
{\large { \bf Supercritical Pomeron \\
and   eikonal representation  \\
 of the diffraction scattering amplitude  }}
\vspace {1.cm}

{\bf S.V.Goloskokov$^1$, S.P.Kuleshov   }\\
Bogoliubov Laboratory of Theoretical Physics,
JINR, Dubna \\
                 and \\
 {\bf          O.V.Selyugin$^{2,3}$} \\
 Division de Physique Th\'{e}orique$^4$, Institut de Physique Nucl\'{e}aire, \\
 91406 Orsay Cedex, France and LPTPE, Universit\'{e} Pierre et Marie Curie, \\
  4 Place Jussieu, 75252 Paris Cedex 05, France \\
\vspace {.5cm}
{\bf Abstract}
\end{center}
   The intercept of the supercritical Pomeron is examined with
the use of different forms of the  scattering amplitudes of the
bare Pomeron.
The one-to-one correspondence between the
eikonal phase and the ratio of the elastic and total cross section is
shown.
   Based on new experimental data of the CDF Collaboration, the
intercept and power of the logarithmic growth of the bare and
total Pomeron amplitude are analyzed.  It is shown that
as a result of the eikonalization procedure, the bare QCD Pomeron
becomes compatible with experiment.
\vspace{1.5cm}
\begin{flushleft}
IPNO/TH 95-12 \hspace{8.5cm} FEBRUARY 1995 \\
------------\\
$^{1} $ E-mail: goloskkv@thsun1.jinr.dubna.su;
$^{2} $ E-mail: selugin@thsun1.jinr.dubna.su\\
$^{3}$ Permanent address: BLThPh, JINR,
141980 Dubna, Russia\\
$^4$ Unit\'{e} de Recherche des Universit\'{e}s Paris 11 et
Paris 6 Associ\'{e}e an CNRS

\end{flushleft}

\newpage
\phantom{.}
\vspace {2.cm}

    Despite a long development of different QCD methods
of the calculations of diffraction processes, the problem of the pomeron
is still very topical. Now we recognize that the research of
the pomeron exchange
requires not only a pure elastic process but also for many
physical processes involving electroweak boson exchanges.
There are two approaches to the pomeron, the "soft" pomeron built of
multiperipheral hadron exchanges and more recent perturbative-QCD
"hard" pomeron built of the gluon-ladder.
     The "soft" pomeron dominates in high energy hadron-hadron
diffractive reactions while the "hard" Pomeron dominates in high energy
$\Upsilon-\Upsilon$ scattering \ci{bj}
and determines the small x-behaviour of deep inelastic structure functions
and spin-averaged gluon distributions.

         The "corner stone" for many models of the Pomeron is the power
of the total cross sections growth.
    In the Regge dynamics, the total cross section in the
high-energy asymptotic
limit is described in terms of a Regge trajectory with intercept
$ \alpha_{pom}(0)=1$.
This leads to a constant total cross section as $s \rightarrow \infty$.
 The growth of the total cross section at ISR energies
requires  new or additional assumptions.
The "soft" pomeron of the standard form with
$ \alpha_{pom}(0)=1 + \epsilon$
was introduced in \ci{lan1}.
  New even and odd under crossing amplitudes were introduced in \ci{nic1}:
the "Froissaron" and "Odderon" which have a maximal power of growth
$\ln^2s/s_0$, where $s_0$ is a scale factor,
and control the growth of cross sections in the asymptotic regime.
 The observed growth of inelastic cross sections and the
multiplicity coincide with these idea.

    Starting from the Low-Nussinov pomeron \ci{lou}, many studies
have been made for determining the pomeron intercept \ci{grave}.
The perturbative QCD leading-log calculation of the gluon ladder diagrams
gives the following result \ci{kur}
$$   \epsilon = 12 \  \frac {\alpha_{s}}{\pi} \  \ln{2} \sim 0.5  $$
This "hard" pomeron is not yet observed experimentally. Really,
the new global QCD analysis of data for various hard scattering processes
leads to the small x behaviour of the gluon structure function
 determined by the "hard" pomeron contribution   \ci{lai}.
$$            g(x) \sim \frac{1}{x^{1+\epsilon}} $$
with $\epsilon = 0.3$.

    So, two different pomeron contributions are observed in soft and hard
processes.  We think that this may be the manifestation of
the same pomeron in different region of momenta transfer.
In hard scattering processes, the interaction time is sufficiently small and
the single "hard" pomeron exchange with $\epsilon \sim 0.3$ contributes.
In soft diffractive hadron reactions, the interaction time is large
and the pomeron rescattering effects must be important.
These contributions can decrease the power of the amplitude growth from
$\epsilon \sim 0.3$ to $\epsilon_{soft} \sim 0.08$.
The same point of view was expressed in \ci{don}.

   A consistent calculation of the rescattering effects is still
an open problem. In what follows, we shall use the most simple
approximation of these  effects, the eikonalization of
the multiple pomeron contributions. It will be shown that the eikonal
representation decreases  $\epsilon \sim 0.14 \div 0.16 $ of the bare pomeron
contribution in the eikonal phase to $\epsilon_{soft} \sim 0.09$.
To perform this analysis, we shall use the ratio of
$\sigma_{el}/\sigma_{tot}$.  It will be shown
that this ratio leads to practically model independent results for
the pomeron intercept.

    There exist many discussions about the energy dependence
of the elastic and total cross section in  hadron - hadron
scattering \ci{blois,mart}.
A recent analysis of the  experimental data has been made
in \ci{block1}. The conclusion has been drawn that this analysis gives
strong evidence for a $\log(s/s_0)$
dependence of $\sigma_{tot}$ at present energies rather than
for $\log^{2}(s/s_0)$ and indicates that the odderon is not
necessary to explain experimental data. In the accordance with this work,
in \ci{jif} the analysis of a phenomenological model has been made
and it has been found that available experimental data at $t=0$ do not
indicate a growth of the total cross sections faster than the first
power in logarithm of the energy.

    On the contrary, in \ci{levin} the overall analysis
of experimental data has been
 made
on the basis of the soft and hard supercritical pomeron where
the $\log^{2}(s/s_0)$ behaviour has been obtained.
Good agreement with the data on deep inelastic scattering
and photoproduction is achieved when the non-perturbative
component of the Pomeron is governed by the "maximal" behaviour, i.e.
 $\ln^{2} s$ \ci{enk2}.

	One can make various comments on this and others analyses
and on the used experimental material. Essential
uncertainty in the values of $\sigma_{tot}$ has been shown in \ci{sel1}.
Now we have a large discussion about the value of $\sigma_{tot}$
at $\sqrt{s} = 1.8 \ TeV$.
In \ci{72} it has been found that at this energy $\sigma_{tot}=72.2 mb$.
Recent results of the CDF Collaboration \ci{CDF}
are
$$ (1+\rho^2) \sigma_{tot}=62.64 \pm 0.95 (mb) \ \ at \ \ \sqrt{s}=546 \ GeV,$$
$$ (1+\rho^2) \sigma_{tot}=81.83 \pm 2.29 (mb) \ \ at \ \ \sqrt{s}=1.8 \ TeV,$$
and
$$ \delta(s_1)= \sigma_{elast} / \sigma_{tot} = 0.210 \pm .002 \ \
at  \ \ \sqrt{s}=546 \ GeV,$$
$$ \delta(s_2)= \sigma_{elast} / \sigma_{tot} = 0.246 \pm .004 \ \
at \ \ \sqrt{s}=1.8 \ TeV.$$

     The last two relations have small errors
becouse  of the cancellation of some errors.
As we will show further, the ratio of these two quantities
is more interesting
and allows us to obtain the intercept of the bare and eikonalized Pomeron.
Let us denote
$$  \Delta(s_{12})= \frac{\delta(s_1)}{\delta(s_2)}=
                  \frac{\sigma_{el}(s_1) \cdot \sigma_{tot}(s_2)}
			 {\sigma_{el}(s_2) \cdot \sigma_{tot}(s_1)}. $$
    We shall consider the properties of the
scattering amplitude at $\sqrt{s} \geq 540 \ GeV$.
At such high energies, we can neglect the contribution of the non-leading
Regge terms, and in this analysis, for simplicity, we neglect the real
part of the scattering amplitude. For the pomeron contribution
we use the ordinary form
\be
T(s,t) = i h s^{\alpha(t)-1} e^{R^2_{0} \cdot t /2} \lab{tt}
\ee
with the linear trajectories $\alpha(t)=\alpha(0)+\alpha^{\prime} t$ and
$\alpha(0)=1 + \varepsilon$.
The differential, elastic and total  cross sections
look as follows:
$$  \frac{d\sigma}{dt}= \pi \mid T(s,t) \mid^{2} ; \ \ \
   \sigma_{el}= \int_{-\infty}^{0} \frac{d\sigma}{dt} dt; \ \ \
   \sigma_{tot}= 4\pi Im T(s,0). $$
The slope $B(s,t)$ at $t=0$ will be defined as
$$      B(s,0) = \frac{d}{dt} [ \ln{(\frac{d\sigma}{dt})}] $$

       Using the scattering amplitude (\ref{tt}) we obtain
\be
      \sigma_{el} = \pi  h^2 \int_{-\infty}^{0} s^{2  (\alpha(t)-1)}
                         e^{R^2_0  t} dt
   \ \	 = \ \  2\pi \frac{ h^2 s^{2 \varepsilon} }{R^2}  \lab{elt}
\ee
and
\be
       \sigma_{tot}= 4\pi h s^{\varepsilon}
\ee
where $R$ can be energy-dependent,
$ R^2 = R^2_{0} (1 + \gamma  ln s) $.

   Hence, the relation $\sigma_{el}/\sigma_{tot}$ is
\be
   \frac{\sigma_{el}}{\sigma_{tot}}= \delta(s) =
    \frac{h}{2 R^2}  s^{\varepsilon}.
\ee
 Using $\Delta(s_{12})$, we can find the intercept
of the pomeron
\be
    \varepsilon= \frac{ ln(\Delta (s_{12}) )}{ln(s_1/s_2)} +
      ln[\frac{1+\gamma ln(s_1)}{1+\gamma ln(s_2)}] / ln(s_1/s_2)  \lab{et}
\ee
$$   =\varepsilon_{0}+\varepsilon_{1}.       $$

      If we  take the amplitude in the logarithmic form
of the energy dependence
\be
T(s,t) = i h \cdot ln^{n}(s) \cdot e^{R^2_{0} \cdot t /2}, \lab{tb}
\ee
we can calculate  $n$
\be
    n= \frac{ ln(\Delta (s_{12}) )}{ln(ln(s_1)/ln(s_2))} +
      ln[\frac{1+\gamma ln(s_1)}{1+\gamma ln(s_2)}]
                        / ln(ln(s_1)/ln(s_2)).    \lab{nt}
\ee
$$	 \ \ \ \      = n_0 + n_1                        $$

   Now let us consider the eikonal representation of the scattering
amplitude which is a simple form of summation of pomeron rescattering
effects:
\be
  T(s,t)= i \  \int \rho d\rho J_{0}(\rho\Delta)(1-e^{i\chi(s,\rho}),
\ee
where
\be
 i \chi(s,\rho)= i \int \Delta J_{0}(\rho\Delta) T^{B}(s,-\Delta^2) d\Delta.
  \lab{gau}
\ee
and $T^{B}$ is the amplitude of the bare pomeron.
Let us use for $T^{B}$ the form (\ref{tt}) and then
for the eikonal phase we have
\be
 i \chi(s,\rho)= - \frac{h s^{\varepsilon^{B}}}{R^2} e^{-\rho^2/(2R^2)} =
 -X \cdot e^{-\rho^2/(2R^2)} .
\ee
It can be shown that the total and
elastic cross section can be represented as
\be
 \sigma_{tot}= - R^2 \ \sum_{k=1}^{\infty} \frac{(-X)^k}{k!k},
\ee

\be
 \sigma_{el}= - R^2 \ \{ 2\sum_{k=1}^{\infty} \frac{(-X)^k}{k!k}
		       + \sum_{k=1}^{\infty}\frac{(-2X)^k}{k!k} \}.
\ee
Hence, their ratio depends only on $X$.

The numerical calculation shows that  for the experimental
values we have the one-to-one correspondence with $X_i$:
\be
 \delta(546)=0.210 \leftrightarrow  X(546)=1.38 \\ \nonumber
 \  \ \ \delta(1800)=0.246 \leftrightarrow  X(1800)=1.862 \lab{xx}
\ee
Based on these values of $X_i$, we can obtain the intercept $\varepsilon^{B}$
or a power of the logarithmic growth $n^{B}$.
They have nearly the same form as  (\ref{et}),(\ref{nt})
\be
    \varepsilon^{B}= \frac{ ln[X(s_1)/X(s_2)]}{ln(s_1/s_2)} +
      ln[\frac{1+\gamma ln(s_1)}{1+\gamma ln(s_2)}] / ln(s_1/s_2)  \lab{eb}
\ee
$$   \ \ \ \    = \varepsilon^{B}_{0} + \varepsilon_{1};   $$

\be
    n^{B}= \frac{ ln[X(s_1)/X(s_2)] }{ln(ln(s_1)/ln(s_2))} +
      ln[\frac{1+\gamma ln(s_1)}{1+\gamma ln(s_2)}] / ln(ln(s_1)/ln(s_2))
\lab{nb}
\ee
$$       =n^{B}_{0}+n_{1}.                                         $$

   Using the recent experimental data we can calculate
$$ \varepsilon_{0}=0.066 \ \ \ \    \varepsilon^{B}_{0}=0.126, $$
$$ n_{0}=0.913, \ \ \ \ n_{0}^{B}=1.73$$.

     The values of $\varepsilon_{1}$ and $n_{1}$ weakly depend
on $\gamma$.
The values of the total cross sections
heavily depend on these values of  $\gamma$.

Taking into account (\ref{xx}) we can calculate  $\sigma_{tot}$ at
$\sqrt{s}= 546$ and $1800 \ GeV$ for different $R_{0}^{2}$.
So, for $R^2=R^2_{0}(1+\gamma ln(s))$ we have , for example:\\

1. for  $\gamma=0$
\begin{tabbing}
$R^2_0=$ \hspace{1.5cm} \= 6.1 \hspace{.5cm}  \= 6.2\hspace{.5cm}
                        \= 6.3 \hspace{.5cm}  \= 6.4 \hspace{.5cm}
                        \= 6.5 \hspace{.5cm}
                                              \\
$\sigma_{tot}(546)=$ \> 60.78 \> 61.77 \> 62.77\> 63.77\> 64.77  \\
$\sigma_{tot}(1800)=$ \>75.03 \> 76.26\> 77.49 \>78.72 \>79.95  \\
\end{tabbing}
and \\

2.    for  $\gamma=0.04$,
\begin{tabbing}
$R^2_0=$   \hspace{1.5cm}  \= 7.8 \hspace{.5cm}     \= 8.0 \hspace{.5cm}   \=
8.2 \\
$\sigma_{tot}(546)=$                  \> 58.45   \> 59.9  \> 61.45 \\
$\sigma_{tot}(1800)=$                 \> 76.74   \> 78.7  \> 80.7
\end{tabbing}

  Let us  recall that the experimental values of $\sigma_{tot}$ are \\
$ (1+ \rho^2 ) \sigma_{tot} (546) = 62.64 \  mb $ CDF \hspace{1cm}
$ (1+\rho^2) \sigma_{tot} (546) =63.5 \ mb $ UA4 \\
$ (1+\rho^2) \sigma_{tot} (1800) =81.83 \ mb $ CDF \hspace{1cm}
and $\sigma_{tot} (546) =63. \ mb $ UA4/2 \\
the last number being a new result of UA4/2 Collaboration \ci{new}.

     It is clear from the above Tables  that the value of $\gamma$
is unlikely to be more than $0.4$ and it is rather preferential
to be near zero
otherwise we will obtain a great divergence from the experimental data
on $\sigma_{tot}$. Moreover, we can conclude that the experimental
value of  $\sigma_{tot}$ at $\sqrt{s} = 1800 \ GeV$
{\it cannot be less} than $78 \div 80 \ mb$;
otherwise it will contradict
either the value of the ratio $\sigma_{el}/\sigma_{tot}$ or the value of
$\sigma_{tot}$ at $\sqrt{s} = 546 \ GeV$.

 The calculation for $\gamma=0.04$ gives
$$ \varepsilon_{1}=0.0257, \ \ \ \  n_{1}=0.35 \ . $$
Hence, as a result, we have for $\gamma=0.04$,
$$ \varepsilon = 0.066+0.026 = 0.092, \ \ \ \
\varepsilon^{B}=0.126+0.026=0.152;$$
$$ n = 0.926+0.35 = 1.28,  \ \ \ \  n^{B} = 1.73 + 0.35 = 2.08. $$

      Now, let us examine a more general case of the
representation of the
eikonal phase \ci{pc}:
\be
 i \chi(s,\rho)= - h(s) \cdot exp(-\mu(s) \cdot \sqrt{\rho^2+b^2(s)})
\ee
 From this representation, one can obtain, in the limit
$b \rightarrow 0$ or $b \rightarrow \infty$ the exponential
or the gaussian form
of the eikonal phase with various  energy dependences
of $h(s)$ and $b(s),\mu(s)$.
This form of the eikonal phase corresponds to
the scattering amplitude
\be
  T(s,0)= -\frac{i}{\mu^{2}} \sum_{n=1}^{\infty}
	\frac{(-X)^n}{n^2 n!} (1+n \mu b);
\ee
here and below $X=h(s) exp(-\mu \cdot b)$ is the eikonal phase at $\rho=0$.
 After calculations one can obtain   $\sigma_{el}$ and $\sigma_{tot}$ :
\be
     \frac{\sigma_{el}}{\sigma_{tot}}=
\frac{1}{2} \frac{[2 T_1(X) -T_1(2X)] + b \cdot \mu [2 T_2(X)-T_2(2X)]}
	      {T_1(X) + b \cdot \mu T_2(X)}    \lab{t1t2}
\ee
Here
\be
 T_1(X)=-\sum_{n=1}^{\infty} \frac{(-X)^{n}}{n^{2} n!}; \, \, \,
 T_2(X)=-\sum_{n=1}^{\infty} \frac{(-X)^{n}}{n n!}.
\ee
    In the case when either $b$ and $ \mu$ do not depend on energy or
\be
  b(s) = b_0 \cdot \kappa(s); \mu(s)=\mu_0 \cdot / \kappa(s)
\ee
the relation (\ref{t1t2}) depends on energy only through $X$.
The calculations show that the value of (\ref{t1t2}) very weakly depends
on the value of $b(s) \cdot \mu(s)$. Again from the value of ratio
$\sigma_{el}(s_i)/\sigma_{tot}(s_i)$ we can obtain the value of $X$
and then calculate  $\sigma_{tot}(s_i)$ and the slope $B(s_i,0)$.

    There is on one more parameters than in the early simplest form
of the eikonal phase (\ref{gau}) so it is needed also to extract
experimental information on the slope $B(s,0)$.
   We can calculate the total cross sections,
the slope $B$ at $\sqrt(s)= 546$ and $1800 \ GeV$
as functions of $\mu_{0}$ and choose
the value of $b_{0}$ so that $\sigma_{tot}(546)=62 \ mb$.
After that we calculate   $\sigma_{tot}$ at $\sqrt{s}=1800 \ GeV$,
the slopes for both energies and the intercept.
These calculations
are shown in Fig.1.
In this figure, the solid lines are the calculated values of
$\sigma_{tot}(1800)$, slopes $B$ at $\sqrt{s}=546 \ GeV$(lower curve) and
at $\sqrt{s}=1800 \ GeV$(upper curve) and intercept of the bare pomeron;
the dot-dashed lines show the experimental values
and the dotted line their errors;
the vertical lines show the bounds on $\mu_{0}$ put by two upper parts
of the figure and hence the values of intercept of the bare pomeron.

   It is clear that the experimental magnitude of
$\sigma_{tot}(1800)=80 \ mb$ restricts weakly the value of $\mu_{0}$,
whereas the magnitudes of slopes at different energies provide strong
bounds.

      We can see that if $\sigma_{tot}$ at $\sqrt{s}=546\ GeV$ is equal
to or larger then  $62 mb$ and we take into account the values of
the relation $\sigma_{el}(s_i)/\sigma_{tot}(s_i)$,
we cannot obtain  $\sigma_{tot}$ at $\sqrt{s}=1800 \ GeV$ less than
$77 \ mb$.

     It is seen from Fig.1 that the intercept
$\epsilon_{bare}= 0.14 \pm 0.01$. The examination of variants with the
energy dependence of $\mu$ and $b$ leads to the same result.


    So we can see that in the examined energy range the power of
the logarithmic growth of $\sigma_{tot}$ is larger than $1$
but
smaller than $2$. It is clear that we have the $ln^2$
term in the total cross section \ci{nicd} but with
a small coefficient, and now we are very far from the asymptotic range.
The ratio $\sigma_{el}/\sigma_{tot} = 0.246$ tells us the same.
It is necessary to note that the analysis performed above clearly shows
that the data obtained for the relations $\sigma_{el}/\sigma_{tot}$ gives
a very strong evidence that  $\sigma_{tot} \simeq 80 \ mb $
at $\sqrt{s}= 1800 \ GeV$. This conclusion is in agreement with one given
already in \ci{nicd}

    In our opinion, this calculation shows that the value
$\varepsilon^{QCD} =0.15 \div 0.17$
 calculated in the framework of the QCD \ci{Braun}
 does not contradict  the phenomenological value
 $ \epsilon = 0.08$.
The $\varepsilon^{QCD}$ is to be  compared with the intercept of the
bare pomeron  $\varepsilon^{B}$, that is  $\sim 0.15$,
as it is evident from our analysis of experimental data.
This is a consequence of the the interaction time being large
in soft diffractive hadron reactions
and we must take into account the pomeron rescattering effects.
Hence the intercept which enters into the structure function can be
sufficiently large in agreement with the value
of the bare pomeron intercept.

\vspace{0.5cm}

     {\it Acknowledgement.} {\hspace {0.5cm} The authors express their deep
gratitude to D.V.Shirkov, A.N. Sissakian and V.A. Meshcheryakov
for support in this work, which  was supported in part by the Russian Fond of
Fundamental Research, Grand $   94-02-04616$. \\
This work was finished during the stay of one of us (O.V.S) at the Institut
de Physique Nucl\'{e}aire, Orsay, France. O.V.Selyugin would like to express
his
warm thanks to D.Vautherin for his kind invitation and B.Nicolescu for
critical remarks.

\newpage

 \begin{picture}(120,220)
 \put(0,50){
\setlength{\unitlength}{0.240900pt}
\ifx\plotpoint\undefined\newsavebox{\plotpoint}\fi
\sbox{\plotpoint}{\rule[-0.200pt]{0.400pt}{0.400pt}}%
\begin{picture}(1500,900)(0,0)
\font\gnuplot=cmr10 at 10pt
\gnuplot
\sbox{\plotpoint}{\rule[-0.200pt]{0.400pt}{0.400pt}}%
\put(220.0,130.0){\rule[-0.200pt]{4.818pt}{0.400pt}}
\put(198,130){\makebox(0,0)[r]{60}}
\put(1416.0,130.0){\rule[-0.200pt]{4.818pt}{0.400pt}}
\put(220.0,286.0){\rule[-0.200pt]{4.818pt}{0.400pt}}
\put(198,286){\makebox(0,0)[r]{65}}
\put(1416.0,286.0){\rule[-0.200pt]{4.818pt}{0.400pt}}
\put(220.0,441.0){\rule[-0.200pt]{4.818pt}{0.400pt}}
\put(198,441){\makebox(0,0)[r]{70}}
\put(1416.0,441.0){\rule[-0.200pt]{4.818pt}{0.400pt}}
\put(220.0,597.0){\rule[-0.200pt]{4.818pt}{0.400pt}}
\put(198,597){\makebox(0,0)[r]{75}}
\put(1416.0,597.0){\rule[-0.200pt]{4.818pt}{0.400pt}}
\put(220.0,753.0){\rule[-0.200pt]{4.818pt}{0.400pt}}
\put(198,753){\makebox(0,0)[r]{80}}
\put(1416.0,753.0){\rule[-0.200pt]{4.818pt}{0.400pt}}
\put(220.0,68.0){\rule[-0.200pt]{0.400pt}{4.818pt}}
\put(220.0,857.0){\rule[-0.200pt]{0.400pt}{4.818pt}}
\put(463.0,68.0){\rule[-0.200pt]{0.400pt}{4.818pt}}
\put(463.0,857.0){\rule[-0.200pt]{0.400pt}{4.818pt}}
\put(706.0,68.0){\rule[-0.200pt]{0.400pt}{4.818pt}}
\put(706.0,857.0){\rule[-0.200pt]{0.400pt}{4.818pt}}
\put(950.0,68.0){\rule[-0.200pt]{0.400pt}{4.818pt}}
\put(950.0,857.0){\rule[-0.200pt]{0.400pt}{4.818pt}}
\put(1193.0,68.0){\rule[-0.200pt]{0.400pt}{4.818pt}}
\put(1193.0,857.0){\rule[-0.200pt]{0.400pt}{4.818pt}}
\put(1436.0,68.0){\rule[-0.200pt]{0.400pt}{4.818pt}}
\put(1436.0,857.0){\rule[-0.200pt]{0.400pt}{4.818pt}}
\put(220.0,68.0){\rule[-0.200pt]{292.934pt}{0.400pt}}
\put(1436.0,68.0){\rule[-0.200pt]{0.400pt}{194.888pt}}
\put(220.0,877.0){\rule[-0.200pt]{292.934pt}{0.400pt}}
\put(45,672){\makebox(0,0){$\sigma_{tot}$ (mb)}}
\put(220.0,68.0){\rule[-0.200pt]{0.400pt}{194.888pt}}
\put(512,877){\usebox{\plotpoint}}
\multiput(512,877)(0.000,-20.756){39}{\usebox{\plotpoint}}
\put(512,68){\usebox{\plotpoint}}
\sbox{\plotpoint}{\rule[-0.500pt]{1.000pt}{1.000pt}}%
\put(220,809){\usebox{\plotpoint}}
\put(220.00,809.00){\usebox{\plotpoint}}
\put(240.76,809.00){\usebox{\plotpoint}}
\multiput(245,809)(20.756,0.000){0}{\usebox{\plotpoint}}
\put(261.51,809.00){\usebox{\plotpoint}}
\multiput(269,809)(20.756,0.000){0}{\usebox{\plotpoint}}
\put(282.27,809.00){\usebox{\plotpoint}}
\put(303.02,809.00){\usebox{\plotpoint}}
\multiput(306,809)(20.756,0.000){0}{\usebox{\plotpoint}}
\put(323.78,809.00){\usebox{\plotpoint}}
\multiput(331,809)(20.756,0.000){0}{\usebox{\plotpoint}}
\put(344.53,809.00){\usebox{\plotpoint}}
\put(365.29,809.00){\usebox{\plotpoint}}
\multiput(367,809)(20.756,0.000){0}{\usebox{\plotpoint}}
\put(386.04,809.00){\usebox{\plotpoint}}
\multiput(392,809)(20.756,0.000){0}{\usebox{\plotpoint}}
\put(406.80,809.00){\usebox{\plotpoint}}
\put(427.55,809.00){\usebox{\plotpoint}}
\multiput(429,809)(20.756,0.000){0}{\usebox{\plotpoint}}
\put(448.31,809.00){\usebox{\plotpoint}}
\multiput(453,809)(20.756,0.000){0}{\usebox{\plotpoint}}
\put(469.07,809.00){\usebox{\plotpoint}}
\put(489.82,809.00){\usebox{\plotpoint}}
\multiput(490,809)(20.756,0.000){0}{\usebox{\plotpoint}}
\put(510.58,809.00){\usebox{\plotpoint}}
\multiput(515,809)(20.756,0.000){0}{\usebox{\plotpoint}}
\put(531.33,809.00){\usebox{\plotpoint}}
\multiput(539,809)(20.756,0.000){0}{\usebox{\plotpoint}}
\put(552.09,809.00){\usebox{\plotpoint}}
\put(572.84,809.00){\usebox{\plotpoint}}
\multiput(576,809)(20.756,0.000){0}{\usebox{\plotpoint}}
\put(593.60,809.00){\usebox{\plotpoint}}
\multiput(601,809)(20.756,0.000){0}{\usebox{\plotpoint}}
\put(614.35,809.00){\usebox{\plotpoint}}
\put(635.11,809.00){\usebox{\plotpoint}}
\multiput(638,809)(20.756,0.000){0}{\usebox{\plotpoint}}
\put(655.87,809.00){\usebox{\plotpoint}}
\multiput(662,809)(20.756,0.000){0}{\usebox{\plotpoint}}
\put(676.62,809.00){\usebox{\plotpoint}}
\put(697.38,809.00){\usebox{\plotpoint}}
\multiput(699,809)(20.756,0.000){0}{\usebox{\plotpoint}}
\put(718.13,809.00){\usebox{\plotpoint}}
\multiput(724,809)(20.756,0.000){0}{\usebox{\plotpoint}}
\put(738.89,809.00){\usebox{\plotpoint}}
\put(759.64,809.00){\usebox{\plotpoint}}
\multiput(760,809)(20.756,0.000){0}{\usebox{\plotpoint}}
\put(780.40,809.00){\usebox{\plotpoint}}
\multiput(785,809)(20.756,0.000){0}{\usebox{\plotpoint}}
\put(801.15,809.00){\usebox{\plotpoint}}
\put(821.91,809.00){\usebox{\plotpoint}}
\multiput(822,809)(20.756,0.000){0}{\usebox{\plotpoint}}
\put(842.66,809.00){\usebox{\plotpoint}}
\multiput(846,809)(20.756,0.000){0}{\usebox{\plotpoint}}
\put(863.42,809.00){\usebox{\plotpoint}}
\multiput(871,809)(20.756,0.000){0}{\usebox{\plotpoint}}
\put(884.18,809.00){\usebox{\plotpoint}}
\put(904.93,809.00){\usebox{\plotpoint}}
\multiput(908,809)(20.756,0.000){0}{\usebox{\plotpoint}}
\put(925.69,809.00){\usebox{\plotpoint}}
\multiput(932,809)(20.756,0.000){0}{\usebox{\plotpoint}}
\put(946.44,809.00){\usebox{\plotpoint}}
\put(967.20,809.00){\usebox{\plotpoint}}
\multiput(969,809)(20.756,0.000){0}{\usebox{\plotpoint}}
\put(987.95,809.00){\usebox{\plotpoint}}
\multiput(994,809)(20.756,0.000){0}{\usebox{\plotpoint}}
\put(1008.71,809.00){\usebox{\plotpoint}}
\put(1029.46,809.00){\usebox{\plotpoint}}
\multiput(1031,809)(20.756,0.000){0}{\usebox{\plotpoint}}
\put(1050.22,809.00){\usebox{\plotpoint}}
\multiput(1055,809)(20.756,0.000){0}{\usebox{\plotpoint}}
\put(1070.98,809.00){\usebox{\plotpoint}}
\put(1091.73,809.00){\usebox{\plotpoint}}
\multiput(1092,809)(20.756,0.000){0}{\usebox{\plotpoint}}
\put(1112.49,809.00){\usebox{\plotpoint}}
\multiput(1117,809)(20.756,0.000){0}{\usebox{\plotpoint}}
\put(1133.24,809.00){\usebox{\plotpoint}}
\multiput(1141,809)(20.756,0.000){0}{\usebox{\plotpoint}}
\put(1154.00,809.00){\usebox{\plotpoint}}
\put(1174.75,809.00){\usebox{\plotpoint}}
\multiput(1178,809)(20.756,0.000){0}{\usebox{\plotpoint}}
\put(1195.51,809.00){\usebox{\plotpoint}}
\multiput(1203,809)(20.756,0.000){0}{\usebox{\plotpoint}}
\put(1216.26,809.00){\usebox{\plotpoint}}
\put(1237.02,809.00){\usebox{\plotpoint}}
\multiput(1239,809)(20.756,0.000){0}{\usebox{\plotpoint}}
\put(1257.77,809.00){\usebox{\plotpoint}}
\multiput(1264,809)(20.756,0.000){0}{\usebox{\plotpoint}}
\put(1278.53,809.00){\usebox{\plotpoint}}
\put(1299.29,809.00){\usebox{\plotpoint}}
\multiput(1301,809)(20.756,0.000){0}{\usebox{\plotpoint}}
\put(1320.04,809.00){\usebox{\plotpoint}}
\multiput(1325,809)(20.756,0.000){0}{\usebox{\plotpoint}}
\put(1340.80,809.00){\usebox{\plotpoint}}
\put(1361.55,809.00){\usebox{\plotpoint}}
\multiput(1362,809)(20.756,0.000){0}{\usebox{\plotpoint}}
\put(1382.31,809.00){\usebox{\plotpoint}}
\multiput(1387,809)(20.756,0.000){0}{\usebox{\plotpoint}}
\put(1403.06,809.00){\usebox{\plotpoint}}
\put(1423.82,809.00){\usebox{\plotpoint}}
\multiput(1424,809)(20.756,0.000){0}{\usebox{\plotpoint}}
\put(1436,809){\usebox{\plotpoint}}
\sbox{\plotpoint}{\rule[-0.200pt]{0.400pt}{0.400pt}}%
\put(220,830){\usebox{\plotpoint}}
\multiput(220.00,828.92)(0.792,-0.497){59}{\rule{0.732pt}{0.120pt}}
\multiput(220.00,829.17)(47.480,-31.000){2}{\rule{0.366pt}{0.400pt}}
\multiput(269.00,797.92)(1.362,-0.497){51}{\rule{1.181pt}{0.120pt}}
\multiput(269.00,798.17)(70.548,-27.000){2}{\rule{0.591pt}{0.400pt}}
\multiput(342.00,770.92)(1.645,-0.498){71}{\rule{1.408pt}{0.120pt}}
\multiput(342.00,771.17)(118.077,-37.000){2}{\rule{0.704pt}{0.400pt}}
\multiput(463.00,733.92)(2.371,-0.497){49}{\rule{1.977pt}{0.120pt}}
\multiput(463.00,734.17)(117.897,-26.000){2}{\rule{0.988pt}{0.400pt}}
\multiput(585.00,707.93)(7.019,-0.489){15}{\rule{5.478pt}{0.118pt}}
\multiput(585.00,708.17)(109.631,-9.000){2}{\rule{2.739pt}{0.400pt}}
\multiput(706.00,698.93)(13.512,-0.477){7}{\rule{9.860pt}{0.115pt}}
\multiput(706.00,699.17)(101.535,-5.000){2}{\rule{4.930pt}{0.400pt}}
\multiput(828.00,693.93)(13.512,-0.477){7}{\rule{9.860pt}{0.115pt}}
\multiput(828.00,694.17)(101.535,-5.000){2}{\rule{4.930pt}{0.400pt}}
\multiput(950.00,688.93)(10.887,-0.482){9}{\rule{8.167pt}{0.116pt}}
\multiput(950.00,689.17)(104.050,-6.000){2}{\rule{4.083pt}{0.400pt}}
\multiput(1071.00,682.93)(10.978,-0.482){9}{\rule{8.233pt}{0.116pt}}
\multiput(1071.00,683.17)(104.911,-6.000){2}{\rule{4.117pt}{0.400pt}}
\multiput(1193.00,676.93)(10.887,-0.482){9}{\rule{8.167pt}{0.116pt}}
\multiput(1193.00,677.17)(104.050,-6.000){2}{\rule{4.083pt}{0.400pt}}
\multiput(1314.00,670.93)(9.265,-0.485){11}{\rule{7.071pt}{0.117pt}}
\multiput(1314.00,671.17)(107.323,-7.000){2}{\rule{3.536pt}{0.400pt}}
\put(220,753){\usebox{\plotpoint}}
\put(220.00,753.00){\usebox{\plotpoint}}
\put(240.76,753.00){\usebox{\plotpoint}}
\multiput(245,753)(20.756,0.000){0}{\usebox{\plotpoint}}
\put(261.51,753.00){\usebox{\plotpoint}}
\multiput(269,753)(20.756,0.000){0}{\usebox{\plotpoint}}
\put(282.27,753.00){\usebox{\plotpoint}}
\put(303.02,753.00){\usebox{\plotpoint}}
\multiput(306,753)(20.756,0.000){0}{\usebox{\plotpoint}}
\put(323.78,753.00){\usebox{\plotpoint}}
\multiput(331,753)(20.756,0.000){0}{\usebox{\plotpoint}}
\put(344.53,753.00){\usebox{\plotpoint}}
\put(365.29,753.00){\usebox{\plotpoint}}
\multiput(367,753)(20.756,0.000){0}{\usebox{\plotpoint}}
\put(386.04,753.00){\usebox{\plotpoint}}
\multiput(392,753)(20.756,0.000){0}{\usebox{\plotpoint}}
\put(406.80,753.00){\usebox{\plotpoint}}
\put(427.55,753.00){\usebox{\plotpoint}}
\multiput(429,753)(20.756,0.000){0}{\usebox{\plotpoint}}
\put(448.31,753.00){\usebox{\plotpoint}}
\multiput(453,753)(20.756,0.000){0}{\usebox{\plotpoint}}
\put(469.07,753.00){\usebox{\plotpoint}}
\put(489.82,753.00){\usebox{\plotpoint}}
\multiput(490,753)(20.756,0.000){0}{\usebox{\plotpoint}}
\put(510.58,753.00){\usebox{\plotpoint}}
\multiput(515,753)(20.756,0.000){0}{\usebox{\plotpoint}}
\put(531.33,753.00){\usebox{\plotpoint}}
\multiput(539,753)(20.756,0.000){0}{\usebox{\plotpoint}}
\put(552.09,753.00){\usebox{\plotpoint}}
\put(572.84,753.00){\usebox{\plotpoint}}
\multiput(576,753)(20.756,0.000){0}{\usebox{\plotpoint}}
\put(593.60,753.00){\usebox{\plotpoint}}
\multiput(601,753)(20.756,0.000){0}{\usebox{\plotpoint}}
\put(614.35,753.00){\usebox{\plotpoint}}
\put(635.11,753.00){\usebox{\plotpoint}}
\multiput(638,753)(20.756,0.000){0}{\usebox{\plotpoint}}
\put(655.87,753.00){\usebox{\plotpoint}}
\multiput(662,753)(20.756,0.000){0}{\usebox{\plotpoint}}
\put(676.62,753.00){\usebox{\plotpoint}}
\put(697.38,753.00){\usebox{\plotpoint}}
\multiput(699,753)(20.756,0.000){0}{\usebox{\plotpoint}}
\put(718.13,753.00){\usebox{\plotpoint}}
\multiput(724,753)(20.756,0.000){0}{\usebox{\plotpoint}}
\put(738.89,753.00){\usebox{\plotpoint}}
\put(759.64,753.00){\usebox{\plotpoint}}
\multiput(760,753)(20.756,0.000){0}{\usebox{\plotpoint}}
\put(780.40,753.00){\usebox{\plotpoint}}
\multiput(785,753)(20.756,0.000){0}{\usebox{\plotpoint}}
\put(801.15,753.00){\usebox{\plotpoint}}
\put(821.91,753.00){\usebox{\plotpoint}}
\multiput(822,753)(20.756,0.000){0}{\usebox{\plotpoint}}
\put(842.66,753.00){\usebox{\plotpoint}}
\multiput(846,753)(20.756,0.000){0}{\usebox{\plotpoint}}
\put(863.42,753.00){\usebox{\plotpoint}}
\multiput(871,753)(20.756,0.000){0}{\usebox{\plotpoint}}
\put(884.18,753.00){\usebox{\plotpoint}}
\put(904.93,753.00){\usebox{\plotpoint}}
\multiput(908,753)(20.756,0.000){0}{\usebox{\plotpoint}}
\put(925.69,753.00){\usebox{\plotpoint}}
\multiput(932,753)(20.756,0.000){0}{\usebox{\plotpoint}}
\put(946.44,753.00){\usebox{\plotpoint}}
\put(967.20,753.00){\usebox{\plotpoint}}
\multiput(969,753)(20.756,0.000){0}{\usebox{\plotpoint}}
\put(987.95,753.00){\usebox{\plotpoint}}
\multiput(994,753)(20.756,0.000){0}{\usebox{\plotpoint}}
\put(1008.71,753.00){\usebox{\plotpoint}}
\put(1029.46,753.00){\usebox{\plotpoint}}
\multiput(1031,753)(20.756,0.000){0}{\usebox{\plotpoint}}
\put(1050.22,753.00){\usebox{\plotpoint}}
\multiput(1055,753)(20.756,0.000){0}{\usebox{\plotpoint}}
\put(1070.98,753.00){\usebox{\plotpoint}}
\put(1091.73,753.00){\usebox{\plotpoint}}
\multiput(1092,753)(20.756,0.000){0}{\usebox{\plotpoint}}
\put(1112.49,753.00){\usebox{\plotpoint}}
\multiput(1117,753)(20.756,0.000){0}{\usebox{\plotpoint}}
\put(1133.24,753.00){\usebox{\plotpoint}}
\multiput(1141,753)(20.756,0.000){0}{\usebox{\plotpoint}}
\put(1154.00,753.00){\usebox{\plotpoint}}
\put(1174.75,753.00){\usebox{\plotpoint}}
\multiput(1178,753)(20.756,0.000){0}{\usebox{\plotpoint}}
\put(1195.51,753.00){\usebox{\plotpoint}}
\multiput(1203,753)(20.756,0.000){0}{\usebox{\plotpoint}}
\put(1216.26,753.00){\usebox{\plotpoint}}
\put(1237.02,753.00){\usebox{\plotpoint}}
\multiput(1239,753)(20.756,0.000){0}{\usebox{\plotpoint}}
\put(1257.77,753.00){\usebox{\plotpoint}}
\multiput(1264,753)(20.756,0.000){0}{\usebox{\plotpoint}}
\put(1278.53,753.00){\usebox{\plotpoint}}
\put(1299.29,753.00){\usebox{\plotpoint}}
\multiput(1301,753)(20.756,0.000){0}{\usebox{\plotpoint}}
\put(1320.04,753.00){\usebox{\plotpoint}}
\multiput(1325,753)(20.756,0.000){0}{\usebox{\plotpoint}}
\put(1340.80,753.00){\usebox{\plotpoint}}
\put(1361.55,753.00){\usebox{\plotpoint}}
\multiput(1362,753)(20.756,0.000){0}{\usebox{\plotpoint}}
\put(1382.31,753.00){\usebox{\plotpoint}}
\multiput(1387,753)(20.756,0.000){0}{\usebox{\plotpoint}}
\put(1403.06,753.00){\usebox{\plotpoint}}
\put(1423.82,753.00){\usebox{\plotpoint}}
\multiput(1424,753)(20.756,0.000){0}{\usebox{\plotpoint}}
\put(1436,753){\usebox{\plotpoint}}
\sbox{\plotpoint}{\rule[-0.500pt]{1.000pt}{1.000pt}}%
\put(220,697){\usebox{\plotpoint}}
\put(220.00,697.00){\usebox{\plotpoint}}
\put(240.76,697.00){\usebox{\plotpoint}}
\multiput(245,697)(20.756,0.000){0}{\usebox{\plotpoint}}
\put(261.51,697.00){\usebox{\plotpoint}}
\multiput(269,697)(20.756,0.000){0}{\usebox{\plotpoint}}
\put(282.27,697.00){\usebox{\plotpoint}}
\put(303.02,697.00){\usebox{\plotpoint}}
\multiput(306,697)(20.756,0.000){0}{\usebox{\plotpoint}}
\put(323.78,697.00){\usebox{\plotpoint}}
\multiput(331,697)(20.756,0.000){0}{\usebox{\plotpoint}}
\put(344.53,697.00){\usebox{\plotpoint}}
\put(365.29,697.00){\usebox{\plotpoint}}
\multiput(367,697)(20.756,0.000){0}{\usebox{\plotpoint}}
\put(386.04,697.00){\usebox{\plotpoint}}
\multiput(392,697)(20.756,0.000){0}{\usebox{\plotpoint}}
\put(406.80,697.00){\usebox{\plotpoint}}
\put(427.55,697.00){\usebox{\plotpoint}}
\multiput(429,697)(20.756,0.000){0}{\usebox{\plotpoint}}
\put(448.31,697.00){\usebox{\plotpoint}}
\multiput(453,697)(20.756,0.000){0}{\usebox{\plotpoint}}
\put(469.07,697.00){\usebox{\plotpoint}}
\put(489.82,697.00){\usebox{\plotpoint}}
\multiput(490,697)(20.756,0.000){0}{\usebox{\plotpoint}}
\put(510.58,697.00){\usebox{\plotpoint}}
\multiput(515,697)(20.756,0.000){0}{\usebox{\plotpoint}}
\put(531.33,697.00){\usebox{\plotpoint}}
\multiput(539,697)(20.756,0.000){0}{\usebox{\plotpoint}}
\put(552.09,697.00){\usebox{\plotpoint}}
\put(572.84,697.00){\usebox{\plotpoint}}
\multiput(576,697)(20.756,0.000){0}{\usebox{\plotpoint}}
\put(593.60,697.00){\usebox{\plotpoint}}
\multiput(601,697)(20.756,0.000){0}{\usebox{\plotpoint}}
\put(614.35,697.00){\usebox{\plotpoint}}
\put(635.11,697.00){\usebox{\plotpoint}}
\multiput(638,697)(20.756,0.000){0}{\usebox{\plotpoint}}
\put(655.87,697.00){\usebox{\plotpoint}}
\multiput(662,697)(20.756,0.000){0}{\usebox{\plotpoint}}
\put(676.62,697.00){\usebox{\plotpoint}}
\put(697.38,697.00){\usebox{\plotpoint}}
\multiput(699,697)(20.756,0.000){0}{\usebox{\plotpoint}}
\put(718.13,697.00){\usebox{\plotpoint}}
\multiput(724,697)(20.756,0.000){0}{\usebox{\plotpoint}}
\put(738.89,697.00){\usebox{\plotpoint}}
\put(759.64,697.00){\usebox{\plotpoint}}
\multiput(760,697)(20.756,0.000){0}{\usebox{\plotpoint}}
\put(780.40,697.00){\usebox{\plotpoint}}
\multiput(785,697)(20.756,0.000){0}{\usebox{\plotpoint}}
\put(801.15,697.00){\usebox{\plotpoint}}
\put(821.91,697.00){\usebox{\plotpoint}}
\multiput(822,697)(20.756,0.000){0}{\usebox{\plotpoint}}
\put(842.66,697.00){\usebox{\plotpoint}}
\multiput(846,697)(20.756,0.000){0}{\usebox{\plotpoint}}
\put(863.42,697.00){\usebox{\plotpoint}}
\multiput(871,697)(20.756,0.000){0}{\usebox{\plotpoint}}
\put(884.18,697.00){\usebox{\plotpoint}}
\put(904.93,697.00){\usebox{\plotpoint}}
\multiput(908,697)(20.756,0.000){0}{\usebox{\plotpoint}}
\put(925.69,697.00){\usebox{\plotpoint}}
\multiput(932,697)(20.756,0.000){0}{\usebox{\plotpoint}}
\put(946.44,697.00){\usebox{\plotpoint}}
\put(967.20,697.00){\usebox{\plotpoint}}
\multiput(969,697)(20.756,0.000){0}{\usebox{\plotpoint}}
\put(987.95,697.00){\usebox{\plotpoint}}
\multiput(994,697)(20.756,0.000){0}{\usebox{\plotpoint}}
\put(1008.71,697.00){\usebox{\plotpoint}}
\put(1029.46,697.00){\usebox{\plotpoint}}
\multiput(1031,697)(20.756,0.000){0}{\usebox{\plotpoint}}
\put(1050.22,697.00){\usebox{\plotpoint}}
\multiput(1055,697)(20.756,0.000){0}{\usebox{\plotpoint}}
\put(1070.98,697.00){\usebox{\plotpoint}}
\put(1091.73,697.00){\usebox{\plotpoint}}
\multiput(1092,697)(20.756,0.000){0}{\usebox{\plotpoint}}
\put(1112.49,697.00){\usebox{\plotpoint}}
\multiput(1117,697)(20.756,0.000){0}{\usebox{\plotpoint}}
\put(1133.24,697.00){\usebox{\plotpoint}}
\multiput(1141,697)(20.756,0.000){0}{\usebox{\plotpoint}}
\put(1154.00,697.00){\usebox{\plotpoint}}
\put(1174.75,697.00){\usebox{\plotpoint}}
\multiput(1178,697)(20.756,0.000){0}{\usebox{\plotpoint}}
\put(1195.51,697.00){\usebox{\plotpoint}}
\multiput(1203,697)(20.756,0.000){0}{\usebox{\plotpoint}}
\put(1216.26,697.00){\usebox{\plotpoint}}
\put(1237.02,697.00){\usebox{\plotpoint}}
\multiput(1239,697)(20.756,0.000){0}{\usebox{\plotpoint}}
\put(1257.77,697.00){\usebox{\plotpoint}}
\multiput(1264,697)(20.756,0.000){0}{\usebox{\plotpoint}}
\put(1278.53,697.00){\usebox{\plotpoint}}
\put(1299.29,697.00){\usebox{\plotpoint}}
\multiput(1301,697)(20.756,0.000){0}{\usebox{\plotpoint}}
\put(1320.04,697.00){\usebox{\plotpoint}}
\multiput(1325,697)(20.756,0.000){0}{\usebox{\plotpoint}}
\put(1340.80,697.00){\usebox{\plotpoint}}
\put(1361.55,697.00){\usebox{\plotpoint}}
\multiput(1362,697)(20.756,0.000){0}{\usebox{\plotpoint}}
\put(1382.31,697.00){\usebox{\plotpoint}}
\multiput(1387,697)(20.756,0.000){0}{\usebox{\plotpoint}}
\put(1403.06,697.00){\usebox{\plotpoint}}
\put(1423.82,697.00){\usebox{\plotpoint}}
\multiput(1424,697)(20.756,0.000){0}{\usebox{\plotpoint}}
\put(1436,697){\usebox{\plotpoint}}
\sbox{\plotpoint}{\rule[-0.200pt]{0.400pt}{0.400pt}}%
\put(220,192){\usebox{\plotpoint}}
\put(220.0,192.0){\rule[-0.200pt]{292.934pt}{0.400pt}}
\put(998,877){\usebox{\plotpoint}}
\multiput(998,877)(0.000,-20.756){39}{\usebox{\plotpoint}}
\put(998,68){\usebox{\plotpoint}}
\end{picture}
}
 \put(0,-155){
\setlength{\unitlength}{0.240900pt}
\ifx\plotpoint\undefined\newsavebox{\plotpoint}\fi
\begin{picture}(1500,900)(0,0)
\font\gnuplot=cmr10 at 10pt
\gnuplot
\sbox{\plotpoint}{\rule[-0.200pt]{0.400pt}{0.400pt}}%
\put(220.0,68.0){\rule[-0.200pt]{4.818pt}{0.400pt}}
\put(198,68){\makebox(0,0)[r]{14.5}}
\put(1416.0,68.0){\rule[-0.200pt]{4.818pt}{0.400pt}}
\put(220.0,270.0){\rule[-0.200pt]{4.818pt}{0.400pt}}
\put(198,270){\makebox(0,0)[r]{15.5}}
\put(1416.0,270.0){\rule[-0.200pt]{4.818pt}{0.400pt}}
\put(220.0,473.0){\rule[-0.200pt]{4.818pt}{0.400pt}}
\put(198,473){\makebox(0,0)[r]{16.5}}
\put(1416.0,473.0){\rule[-0.200pt]{4.818pt}{0.400pt}}
\put(220.0,675.0){\rule[-0.200pt]{4.818pt}{0.400pt}}
\put(198,675){\makebox(0,0)[r]{17.5}}
\put(1416.0,675.0){\rule[-0.200pt]{4.818pt}{0.400pt}}
\put(220.0,877.0){\rule[-0.200pt]{4.818pt}{0.400pt}}
\put(198,877){\makebox(0,0)[r]{18.5}}
\put(1416.0,877.0){\rule[-0.200pt]{4.818pt}{0.400pt}}
\put(220.0,68.0){\rule[-0.200pt]{0.400pt}{4.818pt}}
\put(220.0,857.0){\rule[-0.200pt]{0.400pt}{4.818pt}}
\put(463.0,68.0){\rule[-0.200pt]{0.400pt}{4.818pt}}
\put(463.0,857.0){\rule[-0.200pt]{0.400pt}{4.818pt}}
\put(706.0,68.0){\rule[-0.200pt]{0.400pt}{4.818pt}}
\put(706.0,857.0){\rule[-0.200pt]{0.400pt}{4.818pt}}
\put(950.0,68.0){\rule[-0.200pt]{0.400pt}{4.818pt}}
\put(950.0,857.0){\rule[-0.200pt]{0.400pt}{4.818pt}}
\put(1193.0,68.0){\rule[-0.200pt]{0.400pt}{4.818pt}}
\put(1193.0,857.0){\rule[-0.200pt]{0.400pt}{4.818pt}}
\put(1436.0,68.0){\rule[-0.200pt]{0.400pt}{4.818pt}}
\put(1436.0,857.0){\rule[-0.200pt]{0.400pt}{4.818pt}}
\put(220.0,68.0){\rule[-0.200pt]{292.934pt}{0.400pt}}
\put(1436.0,68.0){\rule[-0.200pt]{0.400pt}{194.888pt}}
\put(220.0,877.0){\rule[-0.200pt]{292.934pt}{0.400pt}}
\put(40,750){\makebox(0,0){$B (GeV^{-2})$}}
\put(220.0,68.0){\rule[-0.200pt]{0.400pt}{194.888pt}}
\put(244,634){\usebox{\plotpoint}}
\multiput(244.58,630.84)(0.498,-0.828){95}{\rule{0.120pt}{0.761pt}}
\multiput(243.17,632.42)(49.000,-79.420){2}{\rule{0.400pt}{0.381pt}}
\multiput(293.58,550.21)(0.498,-0.715){95}{\rule{0.120pt}{0.671pt}}
\multiput(292.17,551.61)(49.000,-68.606){2}{\rule{0.400pt}{0.336pt}}
\multiput(342.00,481.92)(0.510,-0.498){91}{\rule{0.509pt}{0.120pt}}
\multiput(342.00,482.17)(46.945,-47.000){2}{\rule{0.254pt}{0.400pt}}
\multiput(390.00,434.92)(0.831,-0.498){85}{\rule{0.764pt}{0.120pt}}
\multiput(390.00,435.17)(71.415,-44.000){2}{\rule{0.382pt}{0.400pt}}
\multiput(463.00,390.92)(1.073,-0.499){111}{\rule{0.956pt}{0.120pt}}
\multiput(463.00,391.17)(120.015,-57.000){2}{\rule{0.478pt}{0.400pt}}
\multiput(585.00,333.92)(1.190,-0.498){99}{\rule{1.049pt}{0.120pt}}
\multiput(585.00,334.17)(118.823,-51.000){2}{\rule{0.525pt}{0.400pt}}
\multiput(706.00,282.92)(1.393,-0.498){85}{\rule{1.209pt}{0.120pt}}
\multiput(706.00,283.17)(119.490,-44.000){2}{\rule{0.605pt}{0.400pt}}
\multiput(828.00,238.92)(1.705,-0.498){69}{\rule{1.456pt}{0.120pt}}
\multiput(828.00,239.17)(118.979,-36.000){2}{\rule{0.728pt}{0.400pt}}
\multiput(950.00,202.92)(1.740,-0.498){67}{\rule{1.483pt}{0.120pt}}
\multiput(950.00,203.17)(117.922,-35.000){2}{\rule{0.741pt}{0.400pt}}
\multiput(1071.00,167.92)(2.051,-0.497){57}{\rule{1.727pt}{0.120pt}}
\multiput(1071.00,168.17)(118.416,-30.000){2}{\rule{0.863pt}{0.400pt}}
\multiput(1193.00,137.92)(3.071,-0.496){37}{\rule{2.520pt}{0.119pt}}
\multiput(1193.00,138.17)(115.770,-20.000){2}{\rule{1.260pt}{0.400pt}}
\multiput(1314.00,117.92)(5.732,-0.492){19}{\rule{4.536pt}{0.118pt}}
\multiput(1314.00,118.17)(112.585,-11.000){2}{\rule{2.268pt}{0.400pt}}
\sbox{\plotpoint}{\rule[-0.400pt]{0.800pt}{0.800pt}}%
\put(220,442){\usebox{\plotpoint}}
\put(220.0,442.0){\rule[-0.400pt]{292.934pt}{0.800pt}}
\sbox{\plotpoint}{\rule[-0.500pt]{1.000pt}{1.000pt}}%
\put(220,392){\usebox{\plotpoint}}
\put(220.00,392.00){\usebox{\plotpoint}}
\put(240.76,392.00){\usebox{\plotpoint}}
\multiput(245,392)(20.756,0.000){0}{\usebox{\plotpoint}}
\put(261.51,392.00){\usebox{\plotpoint}}
\multiput(269,392)(20.756,0.000){0}{\usebox{\plotpoint}}
\put(282.27,392.00){\usebox{\plotpoint}}
\put(303.02,392.00){\usebox{\plotpoint}}
\multiput(306,392)(20.756,0.000){0}{\usebox{\plotpoint}}
\put(323.78,392.00){\usebox{\plotpoint}}
\multiput(331,392)(20.756,0.000){0}{\usebox{\plotpoint}}
\put(344.53,392.00){\usebox{\plotpoint}}
\put(365.29,392.00){\usebox{\plotpoint}}
\multiput(367,392)(20.756,0.000){0}{\usebox{\plotpoint}}
\put(386.04,392.00){\usebox{\plotpoint}}
\multiput(392,392)(20.756,0.000){0}{\usebox{\plotpoint}}
\put(406.80,392.00){\usebox{\plotpoint}}
\put(427.55,392.00){\usebox{\plotpoint}}
\multiput(429,392)(20.756,0.000){0}{\usebox{\plotpoint}}
\put(448.31,392.00){\usebox{\plotpoint}}
\multiput(453,392)(20.756,0.000){0}{\usebox{\plotpoint}}
\put(469.07,392.00){\usebox{\plotpoint}}
\put(489.82,392.00){\usebox{\plotpoint}}
\multiput(490,392)(20.756,0.000){0}{\usebox{\plotpoint}}
\put(510.58,392.00){\usebox{\plotpoint}}
\multiput(515,392)(20.756,0.000){0}{\usebox{\plotpoint}}
\put(531.33,392.00){\usebox{\plotpoint}}
\multiput(539,392)(20.756,0.000){0}{\usebox{\plotpoint}}
\put(552.09,392.00){\usebox{\plotpoint}}
\put(572.84,392.00){\usebox{\plotpoint}}
\multiput(576,392)(20.756,0.000){0}{\usebox{\plotpoint}}
\put(593.60,392.00){\usebox{\plotpoint}}
\multiput(601,392)(20.756,0.000){0}{\usebox{\plotpoint}}
\put(614.35,392.00){\usebox{\plotpoint}}
\put(635.11,392.00){\usebox{\plotpoint}}
\multiput(638,392)(20.756,0.000){0}{\usebox{\plotpoint}}
\put(655.87,392.00){\usebox{\plotpoint}}
\multiput(662,392)(20.756,0.000){0}{\usebox{\plotpoint}}
\put(676.62,392.00){\usebox{\plotpoint}}
\put(697.38,392.00){\usebox{\plotpoint}}
\multiput(699,392)(20.756,0.000){0}{\usebox{\plotpoint}}
\put(718.13,392.00){\usebox{\plotpoint}}
\multiput(724,392)(20.756,0.000){0}{\usebox{\plotpoint}}
\put(738.89,392.00){\usebox{\plotpoint}}
\put(759.64,392.00){\usebox{\plotpoint}}
\multiput(760,392)(20.756,0.000){0}{\usebox{\plotpoint}}
\put(780.40,392.00){\usebox{\plotpoint}}
\multiput(785,392)(20.756,0.000){0}{\usebox{\plotpoint}}
\put(801.15,392.00){\usebox{\plotpoint}}
\put(821.91,392.00){\usebox{\plotpoint}}
\multiput(822,392)(20.756,0.000){0}{\usebox{\plotpoint}}
\put(842.66,392.00){\usebox{\plotpoint}}
\multiput(846,392)(20.756,0.000){0}{\usebox{\plotpoint}}
\put(863.42,392.00){\usebox{\plotpoint}}
\multiput(871,392)(20.756,0.000){0}{\usebox{\plotpoint}}
\put(884.18,392.00){\usebox{\plotpoint}}
\put(904.93,392.00){\usebox{\plotpoint}}
\multiput(908,392)(20.756,0.000){0}{\usebox{\plotpoint}}
\put(925.69,392.00){\usebox{\plotpoint}}
\multiput(932,392)(20.756,0.000){0}{\usebox{\plotpoint}}
\put(946.44,392.00){\usebox{\plotpoint}}
\put(967.20,392.00){\usebox{\plotpoint}}
\multiput(969,392)(20.756,0.000){0}{\usebox{\plotpoint}}
\put(987.95,392.00){\usebox{\plotpoint}}
\multiput(994,392)(20.756,0.000){0}{\usebox{\plotpoint}}
\put(1008.71,392.00){\usebox{\plotpoint}}
\put(1029.46,392.00){\usebox{\plotpoint}}
\multiput(1031,392)(20.756,0.000){0}{\usebox{\plotpoint}}
\put(1050.22,392.00){\usebox{\plotpoint}}
\multiput(1055,392)(20.756,0.000){0}{\usebox{\plotpoint}}
\put(1070.98,392.00){\usebox{\plotpoint}}
\put(1091.73,392.00){\usebox{\plotpoint}}
\multiput(1092,392)(20.756,0.000){0}{\usebox{\plotpoint}}
\put(1112.49,392.00){\usebox{\plotpoint}}
\multiput(1117,392)(20.756,0.000){0}{\usebox{\plotpoint}}
\put(1133.24,392.00){\usebox{\plotpoint}}
\multiput(1141,392)(20.756,0.000){0}{\usebox{\plotpoint}}
\put(1154.00,392.00){\usebox{\plotpoint}}
\put(1174.75,392.00){\usebox{\plotpoint}}
\multiput(1178,392)(20.756,0.000){0}{\usebox{\plotpoint}}
\put(1195.51,392.00){\usebox{\plotpoint}}
\multiput(1203,392)(20.756,0.000){0}{\usebox{\plotpoint}}
\put(1216.26,392.00){\usebox{\plotpoint}}
\put(1237.02,392.00){\usebox{\plotpoint}}
\multiput(1239,392)(20.756,0.000){0}{\usebox{\plotpoint}}
\put(1257.77,392.00){\usebox{\plotpoint}}
\multiput(1264,392)(20.756,0.000){0}{\usebox{\plotpoint}}
\put(1278.53,392.00){\usebox{\plotpoint}}
\put(1299.29,392.00){\usebox{\plotpoint}}
\multiput(1301,392)(20.756,0.000){0}{\usebox{\plotpoint}}
\put(1320.04,392.00){\usebox{\plotpoint}}
\multiput(1325,392)(20.756,0.000){0}{\usebox{\plotpoint}}
\put(1340.80,392.00){\usebox{\plotpoint}}
\put(1361.55,392.00){\usebox{\plotpoint}}
\multiput(1362,392)(20.756,0.000){0}{\usebox{\plotpoint}}
\put(1382.31,392.00){\usebox{\plotpoint}}
\multiput(1387,392)(20.756,0.000){0}{\usebox{\plotpoint}}
\put(1403.06,392.00){\usebox{\plotpoint}}
\put(1423.82,392.00){\usebox{\plotpoint}}
\multiput(1424,392)(20.756,0.000){0}{\usebox{\plotpoint}}
\put(1436,392){\usebox{\plotpoint}}
\put(220,311){\usebox{\plotpoint}}
\put(220.00,311.00){\usebox{\plotpoint}}
\put(240.76,311.00){\usebox{\plotpoint}}
\multiput(245,311)(20.756,0.000){0}{\usebox{\plotpoint}}
\put(261.51,311.00){\usebox{\plotpoint}}
\multiput(269,311)(20.756,0.000){0}{\usebox{\plotpoint}}
\put(282.27,311.00){\usebox{\plotpoint}}
\put(303.02,311.00){\usebox{\plotpoint}}
\multiput(306,311)(20.756,0.000){0}{\usebox{\plotpoint}}
\put(323.78,311.00){\usebox{\plotpoint}}
\multiput(331,311)(20.756,0.000){0}{\usebox{\plotpoint}}
\put(344.53,311.00){\usebox{\plotpoint}}
\put(365.29,311.00){\usebox{\plotpoint}}
\multiput(367,311)(20.756,0.000){0}{\usebox{\plotpoint}}
\put(386.04,311.00){\usebox{\plotpoint}}
\multiput(392,311)(20.756,0.000){0}{\usebox{\plotpoint}}
\put(406.80,311.00){\usebox{\plotpoint}}
\put(427.55,311.00){\usebox{\plotpoint}}
\multiput(429,311)(20.756,0.000){0}{\usebox{\plotpoint}}
\put(448.31,311.00){\usebox{\plotpoint}}
\multiput(453,311)(20.756,0.000){0}{\usebox{\plotpoint}}
\put(469.07,311.00){\usebox{\plotpoint}}
\put(489.82,311.00){\usebox{\plotpoint}}
\multiput(490,311)(20.756,0.000){0}{\usebox{\plotpoint}}
\put(510.58,311.00){\usebox{\plotpoint}}
\multiput(515,311)(20.756,0.000){0}{\usebox{\plotpoint}}
\put(531.33,311.00){\usebox{\plotpoint}}
\multiput(539,311)(20.756,0.000){0}{\usebox{\plotpoint}}
\put(552.09,311.00){\usebox{\plotpoint}}
\put(572.84,311.00){\usebox{\plotpoint}}
\multiput(576,311)(20.756,0.000){0}{\usebox{\plotpoint}}
\put(593.60,311.00){\usebox{\plotpoint}}
\multiput(601,311)(20.756,0.000){0}{\usebox{\plotpoint}}
\put(614.35,311.00){\usebox{\plotpoint}}
\put(635.11,311.00){\usebox{\plotpoint}}
\multiput(638,311)(20.756,0.000){0}{\usebox{\plotpoint}}
\put(655.87,311.00){\usebox{\plotpoint}}
\multiput(662,311)(20.756,0.000){0}{\usebox{\plotpoint}}
\put(676.62,311.00){\usebox{\plotpoint}}
\put(697.38,311.00){\usebox{\plotpoint}}
\multiput(699,311)(20.756,0.000){0}{\usebox{\plotpoint}}
\put(718.13,311.00){\usebox{\plotpoint}}
\multiput(724,311)(20.756,0.000){0}{\usebox{\plotpoint}}
\put(738.89,311.00){\usebox{\plotpoint}}
\put(759.64,311.00){\usebox{\plotpoint}}
\multiput(760,311)(20.756,0.000){0}{\usebox{\plotpoint}}
\put(780.40,311.00){\usebox{\plotpoint}}
\multiput(785,311)(20.756,0.000){0}{\usebox{\plotpoint}}
\put(801.15,311.00){\usebox{\plotpoint}}
\put(821.91,311.00){\usebox{\plotpoint}}
\multiput(822,311)(20.756,0.000){0}{\usebox{\plotpoint}}
\put(842.66,311.00){\usebox{\plotpoint}}
\multiput(846,311)(20.756,0.000){0}{\usebox{\plotpoint}}
\put(863.42,311.00){\usebox{\plotpoint}}
\multiput(871,311)(20.756,0.000){0}{\usebox{\plotpoint}}
\put(884.18,311.00){\usebox{\plotpoint}}
\put(904.93,311.00){\usebox{\plotpoint}}
\multiput(908,311)(20.756,0.000){0}{\usebox{\plotpoint}}
\put(925.69,311.00){\usebox{\plotpoint}}
\multiput(932,311)(20.756,0.000){0}{\usebox{\plotpoint}}
\put(946.44,311.00){\usebox{\plotpoint}}
\put(967.20,311.00){\usebox{\plotpoint}}
\multiput(969,311)(20.756,0.000){0}{\usebox{\plotpoint}}
\put(987.95,311.00){\usebox{\plotpoint}}
\multiput(994,311)(20.756,0.000){0}{\usebox{\plotpoint}}
\put(1008.71,311.00){\usebox{\plotpoint}}
\put(1029.46,311.00){\usebox{\plotpoint}}
\multiput(1031,311)(20.756,0.000){0}{\usebox{\plotpoint}}
\put(1050.22,311.00){\usebox{\plotpoint}}
\multiput(1055,311)(20.756,0.000){0}{\usebox{\plotpoint}}
\put(1070.98,311.00){\usebox{\plotpoint}}
\put(1091.73,311.00){\usebox{\plotpoint}}
\multiput(1092,311)(20.756,0.000){0}{\usebox{\plotpoint}}
\put(1112.49,311.00){\usebox{\plotpoint}}
\multiput(1117,311)(20.756,0.000){0}{\usebox{\plotpoint}}
\put(1133.24,311.00){\usebox{\plotpoint}}
\multiput(1141,311)(20.756,0.000){0}{\usebox{\plotpoint}}
\put(1154.00,311.00){\usebox{\plotpoint}}
\put(1174.75,311.00){\usebox{\plotpoint}}
\multiput(1178,311)(20.756,0.000){0}{\usebox{\plotpoint}}
\put(1195.51,311.00){\usebox{\plotpoint}}
\multiput(1203,311)(20.756,0.000){0}{\usebox{\plotpoint}}
\put(1216.26,311.00){\usebox{\plotpoint}}
\put(1237.02,311.00){\usebox{\plotpoint}}
\multiput(1239,311)(20.756,0.000){0}{\usebox{\plotpoint}}
\put(1257.77,311.00){\usebox{\plotpoint}}
\multiput(1264,311)(20.756,0.000){0}{\usebox{\plotpoint}}
\put(1278.53,311.00){\usebox{\plotpoint}}
\put(1299.29,311.00){\usebox{\plotpoint}}
\multiput(1301,311)(20.756,0.000){0}{\usebox{\plotpoint}}
\put(1320.04,311.00){\usebox{\plotpoint}}
\multiput(1325,311)(20.756,0.000){0}{\usebox{\plotpoint}}
\put(1340.80,311.00){\usebox{\plotpoint}}
\put(1361.55,311.00){\usebox{\plotpoint}}
\multiput(1362,311)(20.756,0.000){0}{\usebox{\plotpoint}}
\put(1382.31,311.00){\usebox{\plotpoint}}
\multiput(1387,311)(20.756,0.000){0}{\usebox{\plotpoint}}
\put(1403.06,311.00){\usebox{\plotpoint}}
\put(1423.82,311.00){\usebox{\plotpoint}}
\multiput(1424,311)(20.756,0.000){0}{\usebox{\plotpoint}}
\put(1436,311){\usebox{\plotpoint}}
\sbox{\plotpoint}{\rule[-0.400pt]{0.800pt}{0.800pt}}%
\put(220,270){\usebox{\plotpoint}}
\put(220.0,270.0){\rule[-0.400pt]{292.934pt}{0.800pt}}
\sbox{\plotpoint}{\rule[-0.500pt]{1.000pt}{1.000pt}}%
\put(220,230){\usebox{\plotpoint}}
\put(220.00,230.00){\usebox{\plotpoint}}
\put(240.76,230.00){\usebox{\plotpoint}}
\multiput(245,230)(20.756,0.000){0}{\usebox{\plotpoint}}
\put(261.51,230.00){\usebox{\plotpoint}}
\multiput(269,230)(20.756,0.000){0}{\usebox{\plotpoint}}
\put(282.27,230.00){\usebox{\plotpoint}}
\put(303.02,230.00){\usebox{\plotpoint}}
\multiput(306,230)(20.756,0.000){0}{\usebox{\plotpoint}}
\put(323.78,230.00){\usebox{\plotpoint}}
\multiput(331,230)(20.756,0.000){0}{\usebox{\plotpoint}}
\put(344.53,230.00){\usebox{\plotpoint}}
\put(365.29,230.00){\usebox{\plotpoint}}
\multiput(367,230)(20.756,0.000){0}{\usebox{\plotpoint}}
\put(386.04,230.00){\usebox{\plotpoint}}
\multiput(392,230)(20.756,0.000){0}{\usebox{\plotpoint}}
\put(406.80,230.00){\usebox{\plotpoint}}
\put(427.55,230.00){\usebox{\plotpoint}}
\multiput(429,230)(20.756,0.000){0}{\usebox{\plotpoint}}
\put(448.31,230.00){\usebox{\plotpoint}}
\multiput(453,230)(20.756,0.000){0}{\usebox{\plotpoint}}
\put(469.07,230.00){\usebox{\plotpoint}}
\put(489.82,230.00){\usebox{\plotpoint}}
\multiput(490,230)(20.756,0.000){0}{\usebox{\plotpoint}}
\put(510.58,230.00){\usebox{\plotpoint}}
\multiput(515,230)(20.756,0.000){0}{\usebox{\plotpoint}}
\put(531.33,230.00){\usebox{\plotpoint}}
\multiput(539,230)(20.756,0.000){0}{\usebox{\plotpoint}}
\put(552.09,230.00){\usebox{\plotpoint}}
\put(572.84,230.00){\usebox{\plotpoint}}
\multiput(576,230)(20.756,0.000){0}{\usebox{\plotpoint}}
\put(593.60,230.00){\usebox{\plotpoint}}
\multiput(601,230)(20.756,0.000){0}{\usebox{\plotpoint}}
\put(614.35,230.00){\usebox{\plotpoint}}
\put(635.11,230.00){\usebox{\plotpoint}}
\multiput(638,230)(20.756,0.000){0}{\usebox{\plotpoint}}
\put(655.87,230.00){\usebox{\plotpoint}}
\multiput(662,230)(20.756,0.000){0}{\usebox{\plotpoint}}
\put(676.62,230.00){\usebox{\plotpoint}}
\put(697.38,230.00){\usebox{\plotpoint}}
\multiput(699,230)(20.756,0.000){0}{\usebox{\plotpoint}}
\put(718.13,230.00){\usebox{\plotpoint}}
\multiput(724,230)(20.756,0.000){0}{\usebox{\plotpoint}}
\put(738.89,230.00){\usebox{\plotpoint}}
\put(759.64,230.00){\usebox{\plotpoint}}
\multiput(760,230)(20.756,0.000){0}{\usebox{\plotpoint}}
\put(780.40,230.00){\usebox{\plotpoint}}
\multiput(785,230)(20.756,0.000){0}{\usebox{\plotpoint}}
\put(801.15,230.00){\usebox{\plotpoint}}
\put(821.91,230.00){\usebox{\plotpoint}}
\multiput(822,230)(20.756,0.000){0}{\usebox{\plotpoint}}
\put(842.66,230.00){\usebox{\plotpoint}}
\multiput(846,230)(20.756,0.000){0}{\usebox{\plotpoint}}
\put(863.42,230.00){\usebox{\plotpoint}}
\multiput(871,230)(20.756,0.000){0}{\usebox{\plotpoint}}
\put(884.18,230.00){\usebox{\plotpoint}}
\put(904.93,230.00){\usebox{\plotpoint}}
\multiput(908,230)(20.756,0.000){0}{\usebox{\plotpoint}}
\put(925.69,230.00){\usebox{\plotpoint}}
\multiput(932,230)(20.756,0.000){0}{\usebox{\plotpoint}}
\put(946.44,230.00){\usebox{\plotpoint}}
\put(967.20,230.00){\usebox{\plotpoint}}
\multiput(969,230)(20.756,0.000){0}{\usebox{\plotpoint}}
\put(987.95,230.00){\usebox{\plotpoint}}
\multiput(994,230)(20.756,0.000){0}{\usebox{\plotpoint}}
\put(1008.71,230.00){\usebox{\plotpoint}}
\put(1029.46,230.00){\usebox{\plotpoint}}
\multiput(1031,230)(20.756,0.000){0}{\usebox{\plotpoint}}
\put(1050.22,230.00){\usebox{\plotpoint}}
\multiput(1055,230)(20.756,0.000){0}{\usebox{\plotpoint}}
\put(1070.98,230.00){\usebox{\plotpoint}}
\put(1091.73,230.00){\usebox{\plotpoint}}
\multiput(1092,230)(20.756,0.000){0}{\usebox{\plotpoint}}
\put(1112.49,230.00){\usebox{\plotpoint}}
\multiput(1117,230)(20.756,0.000){0}{\usebox{\plotpoint}}
\put(1133.24,230.00){\usebox{\plotpoint}}
\multiput(1141,230)(20.756,0.000){0}{\usebox{\plotpoint}}
\put(1154.00,230.00){\usebox{\plotpoint}}
\put(1174.75,230.00){\usebox{\plotpoint}}
\multiput(1178,230)(20.756,0.000){0}{\usebox{\plotpoint}}
\put(1195.51,230.00){\usebox{\plotpoint}}
\multiput(1203,230)(20.756,0.000){0}{\usebox{\plotpoint}}
\put(1216.26,230.00){\usebox{\plotpoint}}
\put(1237.02,230.00){\usebox{\plotpoint}}
\multiput(1239,230)(20.756,0.000){0}{\usebox{\plotpoint}}
\put(1257.77,230.00){\usebox{\plotpoint}}
\multiput(1264,230)(20.756,0.000){0}{\usebox{\plotpoint}}
\put(1278.53,230.00){\usebox{\plotpoint}}
\put(1299.29,230.00){\usebox{\plotpoint}}
\multiput(1301,230)(20.756,0.000){0}{\usebox{\plotpoint}}
\put(1320.04,230.00){\usebox{\plotpoint}}
\multiput(1325,230)(20.756,0.000){0}{\usebox{\plotpoint}}
\put(1340.80,230.00){\usebox{\plotpoint}}
\put(1361.55,230.00){\usebox{\plotpoint}}
\multiput(1362,230)(20.756,0.000){0}{\usebox{\plotpoint}}
\put(1382.31,230.00){\usebox{\plotpoint}}
\multiput(1387,230)(20.756,0.000){0}{\usebox{\plotpoint}}
\put(1403.06,230.00){\usebox{\plotpoint}}
\put(1423.82,230.00){\usebox{\plotpoint}}
\multiput(1424,230)(20.756,0.000){0}{\usebox{\plotpoint}}
\put(1436,230){\usebox{\plotpoint}}
\sbox{\plotpoint}{\rule[-0.200pt]{0.400pt}{0.400pt}}%
\put(244,877){\usebox{\plotpoint}}
\multiput(244.58,874.53)(0.499,-0.617){193}{\rule{0.120pt}{0.594pt}}
\multiput(243.17,875.77)(98.000,-119.767){2}{\rule{0.400pt}{0.297pt}}
\multiput(342.58,753.77)(0.499,-0.545){239}{\rule{0.120pt}{0.536pt}}
\multiput(341.17,754.89)(121.000,-130.887){2}{\rule{0.400pt}{0.268pt}}
\multiput(463.00,622.92)(0.671,-0.499){179}{\rule{0.636pt}{0.120pt}}
\multiput(463.00,623.17)(120.679,-91.000){2}{\rule{0.318pt}{0.400pt}}
\multiput(585.00,531.92)(1.011,-0.499){117}{\rule{0.907pt}{0.120pt}}
\multiput(585.00,532.17)(119.118,-60.000){2}{\rule{0.453pt}{0.400pt}}
\multiput(706.00,471.92)(1.495,-0.498){79}{\rule{1.290pt}{0.120pt}}
\multiput(706.00,472.17)(119.322,-41.000){2}{\rule{0.645pt}{0.400pt}}
\multiput(828.00,430.92)(1.533,-0.498){77}{\rule{1.320pt}{0.120pt}}
\multiput(828.00,431.17)(119.260,-40.000){2}{\rule{0.660pt}{0.400pt}}
\multiput(950.00,390.92)(1.740,-0.498){67}{\rule{1.483pt}{0.120pt}}
\multiput(950.00,391.17)(117.922,-35.000){2}{\rule{0.741pt}{0.400pt}}
\multiput(1071.00,355.92)(2.809,-0.496){41}{\rule{2.318pt}{0.120pt}}
\multiput(1071.00,356.17)(117.188,-22.000){2}{\rule{1.159pt}{0.400pt}}
\multiput(1193.00,333.92)(4.426,-0.494){25}{\rule{3.557pt}{0.119pt}}
\multiput(1193.00,334.17)(113.617,-14.000){2}{\rule{1.779pt}{0.400pt}}
\multiput(1314.00,319.92)(6.333,-0.491){17}{\rule{4.980pt}{0.118pt}}
\multiput(1314.00,320.17)(111.664,-10.000){2}{\rule{2.490pt}{0.400pt}}
\sbox{\plotpoint}{\rule[-0.500pt]{1.000pt}{1.000pt}}%
\put(220,493){\usebox{\plotpoint}}
\put(220.00,493.00){\usebox{\plotpoint}}
\put(240.76,493.00){\usebox{\plotpoint}}
\multiput(245,493)(20.756,0.000){0}{\usebox{\plotpoint}}
\put(261.51,493.00){\usebox{\plotpoint}}
\multiput(269,493)(20.756,0.000){0}{\usebox{\plotpoint}}
\put(282.27,493.00){\usebox{\plotpoint}}
\put(303.02,493.00){\usebox{\plotpoint}}
\multiput(306,493)(20.756,0.000){0}{\usebox{\plotpoint}}
\put(323.78,493.00){\usebox{\plotpoint}}
\multiput(331,493)(20.756,0.000){0}{\usebox{\plotpoint}}
\put(344.53,493.00){\usebox{\plotpoint}}
\put(365.29,493.00){\usebox{\plotpoint}}
\multiput(367,493)(20.756,0.000){0}{\usebox{\plotpoint}}
\put(386.04,493.00){\usebox{\plotpoint}}
\multiput(392,493)(20.756,0.000){0}{\usebox{\plotpoint}}
\put(406.80,493.00){\usebox{\plotpoint}}
\put(427.55,493.00){\usebox{\plotpoint}}
\multiput(429,493)(20.756,0.000){0}{\usebox{\plotpoint}}
\put(448.31,493.00){\usebox{\plotpoint}}
\multiput(453,493)(20.756,0.000){0}{\usebox{\plotpoint}}
\put(469.07,493.00){\usebox{\plotpoint}}
\put(489.82,493.00){\usebox{\plotpoint}}
\multiput(490,493)(20.756,0.000){0}{\usebox{\plotpoint}}
\put(510.58,493.00){\usebox{\plotpoint}}
\multiput(515,493)(20.756,0.000){0}{\usebox{\plotpoint}}
\put(531.33,493.00){\usebox{\plotpoint}}
\multiput(539,493)(20.756,0.000){0}{\usebox{\plotpoint}}
\put(552.09,493.00){\usebox{\plotpoint}}
\put(572.84,493.00){\usebox{\plotpoint}}
\multiput(576,493)(20.756,0.000){0}{\usebox{\plotpoint}}
\put(593.60,493.00){\usebox{\plotpoint}}
\multiput(601,493)(20.756,0.000){0}{\usebox{\plotpoint}}
\put(614.35,493.00){\usebox{\plotpoint}}
\put(635.11,493.00){\usebox{\plotpoint}}
\multiput(638,493)(20.756,0.000){0}{\usebox{\plotpoint}}
\put(655.87,493.00){\usebox{\plotpoint}}
\multiput(662,493)(20.756,0.000){0}{\usebox{\plotpoint}}
\put(676.62,493.00){\usebox{\plotpoint}}
\put(697.38,493.00){\usebox{\plotpoint}}
\multiput(699,493)(20.756,0.000){0}{\usebox{\plotpoint}}
\put(718.13,493.00){\usebox{\plotpoint}}
\multiput(724,493)(20.756,0.000){0}{\usebox{\plotpoint}}
\put(738.89,493.00){\usebox{\plotpoint}}
\put(759.64,493.00){\usebox{\plotpoint}}
\multiput(760,493)(20.756,0.000){0}{\usebox{\plotpoint}}
\put(780.40,493.00){\usebox{\plotpoint}}
\multiput(785,493)(20.756,0.000){0}{\usebox{\plotpoint}}
\put(801.15,493.00){\usebox{\plotpoint}}
\put(821.91,493.00){\usebox{\plotpoint}}
\multiput(822,493)(20.756,0.000){0}{\usebox{\plotpoint}}
\put(842.66,493.00){\usebox{\plotpoint}}
\multiput(846,493)(20.756,0.000){0}{\usebox{\plotpoint}}
\put(863.42,493.00){\usebox{\plotpoint}}
\multiput(871,493)(20.756,0.000){0}{\usebox{\plotpoint}}
\put(884.18,493.00){\usebox{\plotpoint}}
\put(904.93,493.00){\usebox{\plotpoint}}
\multiput(908,493)(20.756,0.000){0}{\usebox{\plotpoint}}
\put(925.69,493.00){\usebox{\plotpoint}}
\multiput(932,493)(20.756,0.000){0}{\usebox{\plotpoint}}
\put(946.44,493.00){\usebox{\plotpoint}}
\put(967.20,493.00){\usebox{\plotpoint}}
\multiput(969,493)(20.756,0.000){0}{\usebox{\plotpoint}}
\put(987.95,493.00){\usebox{\plotpoint}}
\multiput(994,493)(20.756,0.000){0}{\usebox{\plotpoint}}
\put(1008.71,493.00){\usebox{\plotpoint}}
\put(1029.46,493.00){\usebox{\plotpoint}}
\multiput(1031,493)(20.756,0.000){0}{\usebox{\plotpoint}}
\put(1050.22,493.00){\usebox{\plotpoint}}
\multiput(1055,493)(20.756,0.000){0}{\usebox{\plotpoint}}
\put(1070.98,493.00){\usebox{\plotpoint}}
\put(1091.73,493.00){\usebox{\plotpoint}}
\multiput(1092,493)(20.756,0.000){0}{\usebox{\plotpoint}}
\put(1112.49,493.00){\usebox{\plotpoint}}
\multiput(1117,493)(20.756,0.000){0}{\usebox{\plotpoint}}
\put(1133.24,493.00){\usebox{\plotpoint}}
\multiput(1141,493)(20.756,0.000){0}{\usebox{\plotpoint}}
\put(1154.00,493.00){\usebox{\plotpoint}}
\put(1174.75,493.00){\usebox{\plotpoint}}
\multiput(1178,493)(20.756,0.000){0}{\usebox{\plotpoint}}
\put(1195.51,493.00){\usebox{\plotpoint}}
\multiput(1203,493)(20.756,0.000){0}{\usebox{\plotpoint}}
\put(1216.26,493.00){\usebox{\plotpoint}}
\put(1237.02,493.00){\usebox{\plotpoint}}
\multiput(1239,493)(20.756,0.000){0}{\usebox{\plotpoint}}
\put(1257.77,493.00){\usebox{\plotpoint}}
\multiput(1264,493)(20.756,0.000){0}{\usebox{\plotpoint}}
\put(1278.53,493.00){\usebox{\plotpoint}}
\put(1299.29,493.00){\usebox{\plotpoint}}
\multiput(1301,493)(20.756,0.000){0}{\usebox{\plotpoint}}
\put(1320.04,493.00){\usebox{\plotpoint}}
\multiput(1325,493)(20.756,0.000){0}{\usebox{\plotpoint}}
\put(1340.80,493.00){\usebox{\plotpoint}}
\put(1361.55,493.00){\usebox{\plotpoint}}
\multiput(1362,493)(20.756,0.000){0}{\usebox{\plotpoint}}
\put(1382.31,493.00){\usebox{\plotpoint}}
\multiput(1387,493)(20.756,0.000){0}{\usebox{\plotpoint}}
\put(1403.06,493.00){\usebox{\plotpoint}}
\put(1423.82,493.00){\usebox{\plotpoint}}
\multiput(1424,493)(20.756,0.000){0}{\usebox{\plotpoint}}
\put(1436,493){\usebox{\plotpoint}}
\sbox{\plotpoint}{\rule[-0.200pt]{0.400pt}{0.400pt}}%
\put(512,877){\usebox{\plotpoint}}
\multiput(512,877)(0.000,-20.756){39}{\usebox{\plotpoint}}
\put(512,68){\usebox{\plotpoint}}
\put(998,877){\usebox{\plotpoint}}
\multiput(998,877)(0.000,-20.756){39}{\usebox{\plotpoint}}
\put(998,68){\usebox{\plotpoint}}
\end{picture}
}
 \put(0,-360){
\setlength{\unitlength}{0.240900pt}
\ifx\plotpoint\undefined\newsavebox{\plotpoint}\fi
\sbox{\plotpoint}{\rule[-0.200pt]{0.400pt}{0.400pt}}%
\begin{picture}(1500,900)(0,0)
\font\gnuplot=cmr10 at 10pt
\gnuplot
\sbox{\plotpoint}{\rule[-0.200pt]{0.400pt}{0.400pt}}%
\put(220.0,68.0){\rule[-0.200pt]{4.818pt}{0.400pt}}
\put(198,68){\makebox(0,0)[r]{0.1}}
\put(1416.0,68.0){\rule[-0.200pt]{4.818pt}{0.400pt}}
\put(220.0,270.0){\rule[-0.200pt]{4.818pt}{0.400pt}}
\put(198,270){\makebox(0,0)[r]{0.12}}
\put(1416.0,270.0){\rule[-0.200pt]{4.818pt}{0.400pt}}
\put(220.0,472.0){\rule[-0.200pt]{4.818pt}{0.400pt}}
\put(198,472){\makebox(0,0)[r]{0.14}}
\put(1416.0,472.0){\rule[-0.200pt]{4.818pt}{0.400pt}}
\put(220.0,675.0){\rule[-0.200pt]{4.818pt}{0.400pt}}
\put(198,675){\makebox(0,0)[r]{0.16}}
\put(1416.0,675.0){\rule[-0.200pt]{4.818pt}{0.400pt}}
\put(220.0,877.0){\rule[-0.200pt]{4.818pt}{0.400pt}}
\put(198,877){\makebox(0,0)[r]{0.18}}
\put(1416.0,877.0){\rule[-0.200pt]{4.818pt}{0.400pt}}
\put(220.0,68.0){\rule[-0.200pt]{0.400pt}{4.818pt}}
\put(220,23){\makebox(0,0){0.5}}
\put(220.0,857.0){\rule[-0.200pt]{0.400pt}{4.818pt}}
\put(463.0,68.0){\rule[-0.200pt]{0.400pt}{4.818pt}}
\put(463,23){\makebox(0,0){0.6}}
\put(463.0,857.0){\rule[-0.200pt]{0.400pt}{4.818pt}}
\put(706.0,68.0){\rule[-0.200pt]{0.400pt}{4.818pt}}
\put(706,23){\makebox(0,0){0.7}}
\put(706.0,857.0){\rule[-0.200pt]{0.400pt}{4.818pt}}
\put(950.0,68.0){\rule[-0.200pt]{0.400pt}{4.818pt}}
\put(950,23){\makebox(0,0){0.8}}
\put(950.0,857.0){\rule[-0.200pt]{0.400pt}{4.818pt}}
\put(1193.0,68.0){\rule[-0.200pt]{0.400pt}{4.818pt}}
\put(1193,23){\makebox(0,0){0.9}}
\put(1193.0,857.0){\rule[-0.200pt]{0.400pt}{4.818pt}}
\put(1436.0,68.0){\rule[-0.200pt]{0.400pt}{4.818pt}}
\put(1436,23){\makebox(0,0){1}}
\put(1350,-10){\makebox(0,0){$\mu_{0}$ \ GeV}}
\put(1436.0,857.0){\rule[-0.200pt]{0.400pt}{4.818pt}}
\put(220.0,68.0){\rule[-0.200pt]{292.934pt}{0.400pt}}
\put(1436.0,68.0){\rule[-0.200pt]{0.400pt}{194.888pt}}
\put(220.0,877.0){\rule[-0.200pt]{292.934pt}{0.400pt}}
\put(55,750){\makebox(0,0){{\large $\varepsilon_{bare}$}}}
\put(220.0,68.0){\rule[-0.200pt]{0.400pt}{194.888pt}}
\put(220,852){\usebox{\plotpoint}}
\multiput(220.58,849.31)(0.499,-0.685){241}{\rule{0.120pt}{0.648pt}}
\multiput(219.17,850.66)(122.000,-165.656){2}{\rule{0.400pt}{0.324pt}}
\multiput(342.00,683.92)(0.545,-0.499){219}{\rule{0.536pt}{0.120pt}}
\multiput(342.00,684.17)(119.887,-111.000){2}{\rule{0.268pt}{0.400pt}}
\multiput(463.00,572.92)(0.860,-0.499){139}{\rule{0.787pt}{0.120pt}}
\multiput(463.00,573.17)(120.366,-71.000){2}{\rule{0.394pt}{0.400pt}}
\multiput(585.00,501.92)(1.691,-0.498){69}{\rule{1.444pt}{0.120pt}}
\multiput(585.00,502.17)(118.002,-36.000){2}{\rule{0.722pt}{0.400pt}}
\multiput(706.00,465.92)(3.096,-0.496){37}{\rule{2.540pt}{0.119pt}}
\multiput(706.00,466.17)(116.728,-20.000){2}{\rule{1.270pt}{0.400pt}}
\multiput(828.00,445.92)(2.051,-0.497){57}{\rule{1.727pt}{0.120pt}}
\multiput(828.00,446.17)(118.416,-30.000){2}{\rule{0.863pt}{0.400pt}}
\multiput(950.00,415.92)(4.123,-0.494){27}{\rule{3.327pt}{0.119pt}}
\multiput(950.00,416.17)(114.095,-15.000){2}{\rule{1.663pt}{0.400pt}}
\multiput(1071.00,400.92)(6.333,-0.491){17}{\rule{4.980pt}{0.118pt}}
\multiput(1071.00,401.17)(111.664,-10.000){2}{\rule{2.490pt}{0.400pt}}
\multiput(1193.00,390.92)(5.684,-0.492){19}{\rule{4.500pt}{0.118pt}}
\multiput(1193.00,391.17)(111.660,-11.000){2}{\rule{2.250pt}{0.400pt}}
\multiput(1314.00,379.92)(6.333,-0.491){17}{\rule{4.980pt}{0.118pt}}
\multiput(1314.00,380.17)(111.664,-10.000){2}{\rule{2.490pt}{0.400pt}}
\put(512,68){\usebox{\plotpoint}}
\multiput(512,68)(0.000,20.756){35}{\usebox{\plotpoint}}
\put(512,776){\usebox{\plotpoint}}
\put(998,68){\usebox{\plotpoint}}
\multiput(998,68)(0.000,20.756){35}{\usebox{\plotpoint}}
\put(998,776){\usebox{\plotpoint}}
\end{picture}
}
 \put(0,-385){
{\bf Fig.1} The dependence of $\sigma_{tot}(s_i)$, slope-$B(0,s_i)$
and intercept of bare Pomeron
on $\mu_{0}$ .
}
\end{picture}

\newpage

\phantom{.}
\vspace {2cm}


{\bf Figure Captions}

\phantom{.} \hspace{1cm} {\bf Fig.1} The dependence of
$\sigma_{tot}(s_i)$, slope-$B(0,s_i)$ and intercept of bare Pomeron
on $\mu_{0}$ .

\end{document}